\documentclass[prd,superscriptaddress,altaffilletter,showpacs,nofootinbib,twocolumn]{revtex4}

\usepackage[T1]{fontenc}    
\usepackage{hyperref}       
\usepackage{url}            
\usepackage{booktabs}       
\usepackage{amsfonts}       
\usepackage{nicefrac}       
\usepackage{microtype}      
\usepackage{lipsum}
\usepackage{graphicx}
\usepackage[T1]{fontenc} 
\usepackage{graphicx,color}
\usepackage[dvipsnames]{xcolor}
\usepackage{soul}
\usepackage{url}
\usepackage{subcaption}
\usepackage{amssymb}
\usepackage{float}
\usepackage{pdf14}
\usepackage{tikz}
\usepackage{placeins}
\usepackage{algorithm2e}
\usepackage{algorithmic}
\usepackage{amsmath}

\definecolor{owngreen}{rgb}{0.0, 0.5, 0.0}

\newcommand{\Like}{\mathcal{L}}
\usepackage{color}

\graphicspath{ {./img/} }
\newcommand{\changes}[1]{\textcolor{black}{#1}}

\begin{document}


\title{Deep Learning and genetic algorithms for cosmological Bayesian inference speed-up}

\author{Isidro G\'omez-Vargas}\email{igomez@icf.unam.mx}\affiliation{Instituto de Ciencias F\'isicas, Universidad Nacional Aut\'onoma de M\'exico, 62210, Cuernavaca, Morelos, M\'exico.}
\affiliation{Department of Astronomy, University of Geneva, Versoix, 1290, Switzerland.}

\author{J. Alberto V\'azquez }\email{javazquez@icf.unam.mx}\affiliation{Instituto de Ciencias F\'isicas, Universidad Nacional Aut\'onoma de M\'exico, 62210, Cuernavaca, Morelos, M\'exico.}

\date{\today}

\begin{abstract}
In this paper, we present a novel approach to accelerate the Bayesian inference process, focusing specifically on the nested sampling algorithms. Bayesian inference plays a crucial role in cosmological parameter estimation, providing a robust framework for extracting theoretical insights from observational data. However, its computational demands can be substantial, primarily due to the need for numerous likelihood function evaluations.
Our method utilizes the power of deep learning, employing feedforward neural networks to approximate the likelihood function dynamically during the Bayesian inference process. Unlike traditional approaches, our method trains neural networks on-the-fly using the current set of live points as training data, without the need for pre-training. This flexibility enables adaptation to various theoretical models and datasets.
We perform the hyperparameter optimization using genetic algorithms to suggest initial neural network architectures for learning each likelihood function. Once sufficient accuracy is achieved, the neural network replaces the original likelihood function. The implementation integrates with nested sampling algorithms and has been thoroughly evaluated using both simple cosmological dark energy models and diverse observational datasets.
Additionally, we explore the potential of genetic algorithms for generating initial live points within nested sampling inference, opening up new avenues for enhancing the efficiency and effectiveness of Bayesian inference methods.
\end{abstract}


\maketitle
%
\section{Introduction}
\label{sec:intro}
Bayesian inference is a powerful tool in several scientific fields where it is essential to constrain mathematical models using experimental data. It allows parameter estimation and model comparison. In particular, it is the data analysis technique per excellence in observational cosmology, as it provides a robust method to obtain valuable statistical information from a theoretical model given a set of observational data. However, a significant disadvantage of Bayesian inference lies in its high computational cost; it requires a considerable number of likelihood function evaluations to generate sufficient samples from the posterior distribution. For example, a small Bayesian inference task could involve thousands of samples and require thousands, or even millions, of likelihood evaluations.

Given the crucial importance of parameter estimation in the context of astronomical surveys, within the fields of cosmology and astrophysics, numerous valuable efforts have been made to address the computational challenge of mitigating the complexity of the likelihood function calculation to speed up Bayesian inference. Some strategies provide an approximation of Bayesian inference by avoiding the computation of the full likelihood function, as suggested by \cite{akeret2015approximate, jennings2017astroabc, ISHIDA20151}. On the other hand, some efforts try to speed up the inference with different statistical techniques \cite{petrosyan2023supernest, dunkley2005, brinckmann2019montepython, schuhmann2016gaussianization, lewis2013efficient}. Alternatively, other works \cite{sato2011copula, fendt2007pico, pellejero2020cosmological, alsing2019fast} introduced the concept of generating synthetic likelihood distributions. Furthermore, there is an emerging trend of exploiting machine learning tools to accelerate the Bayesian inference process \cite{fendt2007pico, moss2020accelerated, hortua2020accelerating, gomez2021neural, spurio2022cosmopower}.

The use of artificial neural networks (ANNs) to approximate the likelihood function can greatly improve the efficiency of Bayesian inference \cite{auld2007fast, graff2012, graff2014, hortua2020accelerating, hortua2020parameter, spurio2022cosmopower, nygaard2023connect}.  However, it is necessary to have a careful consideration of the trade-off between accuracy and speed, along with quality monitoring of the resulting posterior samples. In addition, neural networks present several drawbacks that must be taken into account to effectively aid in the performance of Bayesian inference:

\begin{enumerate}
    \item \textbf{ANNs excel at interpolation, but not at extrapolation.} Like all machine learning algorithms, ANNs generate models based on datasets, allowing them to learn data structures and predict unseen data within the bounds of the training region. In the Bayesian inference domain, new samples try to find better likelihood values, which could correspond to points outside the ranges of the random sample used for the ANN training.
    \item \textbf{The performance of ANNs depends on their hyperparameters}. This is perhaps one of the most challenging issues facing neural networks. If the hyperparameters are not chosen carefully, the neural network models can be under- or over-fitted.
    \item \textbf{The selection of hyperparameters depends on the data}. There is no unique architecture for an ANN. Each dataset requires certain hyperparameter configurations to have an efficient training of the neural network.
    \item \textbf{Training an ANN requires computational resources}. It is a well-known fact that training a neural network can be computationally demanding, which seems contradictory when the goal is to reduce the computational time in a Bayesian inference process.
\end{enumerate}

We will come back to these issues in Section \ref{sec:strategies} by presenting how each of them is addressed by the method we propose.  

Previous works using neural networks in cosmological parameter estimation save an amazing amount of computational time training neural networks before the Bayesian inference process \cite{spurio2022cosmopower, hortua2020accelerating, chantada2023nn}; however, the pre-training time in these cases is expensive and the trained neural networks are only useful for a specific configuration of backgrounds, models, and data sets. For this reason, our work is inspired by BAMBI \cite{graff2012, graff2014}, and pyBambi \cite{pybambi}, where their neural networks are trained in real-time to learn the likelihood function, which is subsequently replaced within a nested sampling process. The strength of this approach lies in its ability to train the neural network in real-time and accelerate the Bayesian inference process without restricting a cosmological or theoretical model and specific datasets.  In our method, we explore features beyond those of our predecessors, such as parallelism, PyTorch implementation \cite{paszke2019pytorch}, and hyperparameter tuning. In addition, we exclusively used live points for training to reduce the dispersion of the training dataset and to obtain results with higher accuracy. A criterion was also chosen to initiate our method that serves as a regulator of the trade-off between accuracy and speed. We also implemented an on-the-fly performance evaluation to accept or reject the neural network predictions. In addition, we have conducted a preliminary investigation on the use of genetic algorithms to generate the initial sample of live points on the nested sampling process.
\\

The structure of the paper is as follows: Section \ref{sec:background} offers an overview of Bayesian inference and nested sampling. Section \ref{sec:ML_background} provides a concise exposition of the machine learning fundamentals employed in this study. The concept and development of our machine learning strategies are detailed in Section \ref{sec:strategies}. Section \ref{sec:toymodels} and Section \ref{sec:cosmo} present our results, applied respectively to testing toy models and estimating cosmological parameters. In Section \ref{sec:conclusions}, we discuss our research findings and present our final reflections. Furthermore, the Appendix features preliminary results about the incorporation of genetic algorithms as initiators of the live points in a nested sampling execution.

\section{Statistical background}

\label{sec:background}
In this section, we describe an overview of Bayesian inference and neural networks. In particular, we focus on the nested sampling algorithm and feedforward neural networks.
\subsection{Bayesian inference}
\label{sec:bayesian}
    Considering the Bayes' theorem as follows:
        \begin{equation}
            P(\theta| D)= \frac{P(D|\theta)P(\theta)}{P(D)},
            \label{eq:bayesbayesian}
        \end{equation}
    where $P(\theta)$ denotes the prior distribution over parameters $\theta$, encapsulating any prior knowledge about them before observing the data. $P(D|\theta)$ represents the likelihood function, expressing the conditional probability of observing the data given the model. Finally,  the Bayesian evidence $P(D)$ serves as a normalization constant through likelihood marginalization: 
    \begin{equation}
        P(D)=\int^N_\theta  P(D|\theta)P(\theta)d\theta,
        \label{eq:evidence}
    \end{equation}
    where $N$ is the number of dimensions of the parameter space for $\theta$. 
    
    It can be assumed that the measurement error $\epsilon$ is independent of $\theta$ and has a Probability Density Function (PDF) $P_\epsilon$. In this case, the predicted value and the measurement error share the same distribution, therefore the likelihood function can be expressed as:
        \begin{equation}
            P(D|\theta) = P_\epsilon(D - f(x;\theta)),
            \label{eq:like1}
        \end{equation}
    and if the error $\epsilon \sim N(0, C)$ has a normal distribution centered in zero and a covariance matrix $C$, then we have the following: 
        \begin{equation}
        P(D|\theta) = \frac{1}{(2 \pi)^{N/2}|C|^{1/2}} e^{-0.5(D-f(x;\theta))^T C^{-1}(D-f(x;\theta))}  \;\;.  
        \label{eq:like2}
        \end{equation}

   \subsection{Nested sampling}
     Nested sampling (NS) belongs to a category of inference methods that estimate the Bayesian evidence along with its uncertainty by sampling the posterior probability density function. It was proposed by John Skilling in 2004 \cite{skilling2004nested, skilling2006nested}. The evidence, or marginalization of the likelihood function, is a key quantity in model comparison, through the Bayes factor in the Jeffreys’ scale. It is a more rigorous technique \cite{raftery1996approximate, skilling2006nested} than other widely used methods such as the information criteria approximations \cite{liddle2007information, liddle2006cosmological}. NS works by computing the Bayesian evidence while assuming that the parameter space (prior volume or prior mass) shrinks by a certain factor. 
     There are successful Nested Sampling implementations \cite{feroz2009multinest, handley2015, speagle2018} and several applications in cosmology \cite{trotta2013recent, parkinson2011cosmonest, mukherjee2006nested, audren2013conservative, akrami2020planck}, astrophysics \cite{bernst2011magix, buchner2014x, corsaro2014diamonds}, gravitational waves analysis \cite{feroz2009use, pitkin2012new, del2011testing}, biology \cite{pullen2014bayesian, aitken2013nested} and in other scientific fields \cite{partay2014nested, baldock2017classical, szekeres2018direct}.
        
     To understand the method proposed in this work, we briefly describe some considerations about the NS algorithm. For more details, we recommend Refs. \cite{skilling2006nested, handley2015, speagle2018}. First of all, the Bayesian evidence can be written as follows:
    
      \begin{equation}
            Z = \int \Like (\theta) \pi (\theta) d\theta ,
            \label{eq:bayesian_evidence}
        \end{equation}
    where $\theta$ represents the free parameters, $\pi(\theta)$ is the prior density, and $\Like$ is the likelihood function.
    
    The basic idea of NS is to simplify the integration of Bayesian evidence by mapping the parameter space in a unit hypercube. The fraction of the prior contained within an iso-likelihood contour $\Like_c$ in the unit hypercube is called prior volume (or prior mass):
    
        \begin{equation}
            X(\Like) = \int_{\Like(\theta)>\Like_c} \pi(\theta)d\theta .
            \label{eq:priorvol}
        \end{equation}
    The Bayesian evidence can be reduced as a one-dimensional integral of the Likelihood as a function of the prior volume $X$:
            
        \begin{equation}
            Z = \int_0^1 \Like(X) dX. 
            \label{eq:evidence1d}
        \end{equation}
    
    NS starts with a specific number $n_{\rm live}$ of random points, termed live points, distributed within the prior volume defined by the constrained prior. These samples are ordered based on their likelihood values. During each iteration the worst point $\Like_{\rm worst}$, with the lowest likelihood value, is removed. A new sample is then generated within a contour bounded by $\Like_{\rm worst}$ and with a likelihood, $\Like(\theta) > \Like_{\rm worst}$. Equation \ref{eq:evidence1d} can be simplified as a Riemann sum:
    \begin{equation}
        Z \approx \sum^N_{i=1} L_i\omega_i ,
        \label{eq:zriemmansum}
    \end{equation}
     where $\omega_i$ is the difference between the prior volume of two consecutive points: $\omega_i = X_{i-1} - X_i$. Throughout the process, NS retains the population of $n_{\rm live}$ live points and ultimately consolidates the final set of live points within a region of high probability. Depending on the sampling approach employed from the constrained prior, various nested sampling algorithms exist. For instance, \texttt{MultiNest} \cite{feroz2009multinest} utilizes rejection sampling within ellipsoids, whereas \texttt{Polychord} \cite{handley2015} generates points using slice sampling.
          
    Several stopping criteria exist for terminating a nested sampling run; in this study, we adopt the remaining evidence criterion, which is roughly outlined as follows:
    \begin{equation}
        \Delta Z_i \approx \Like_{\rm max}X_i,
    \label{eq:dlogz}
    \end{equation}
    hence defining the logarithmic ratio between the current estimated evidence and the remaining evidence as:
    \begin{equation}
        \Delta \ln Z_i \equiv \ln(Z_i + \Delta Z_i) - \ln Z_i,
    \end{equation}
    referred to as \texttt{dlogz} hereafter in this paper. Stopping at a value \texttt{dlogz} implies sampling until only a fraction of the evidence remains unaccounted for.
    
\section{Machine Learning background}
\label{sec:ML_background}
Machine learning is the field of Artificial Intelligence concerning to the mathematical modeling of datasets. Its methods identify inherent properties of datasets by minimizing a target function until it reaches a satisfactory value. Over the past few years, Artificial Neural Networks (ANNs) have emerged as the most successful type of machine learning models, giving rise to the field of deep learning. On the other hand, genetic algorithms are a special class of evolutionary algorithms, called metaheuristics, facilitating function optimization without derivatives.

This section offers a succinct overview of artificial neural networks and genetic algorithms.

\subsection{Artificial neural networks}    
\label{sec:ann}
An artificial neural network (ANN) is a computational model inspired by biological synapses, aiming to replicate their behavior. It consists of interconnected layers of nodes, or neurons, serving as basic processing units. A fundamental type of ANN is the feedforward neural network, comprising input, hidden, and output layers. In such networks, connections between neurons, known as weights, are parameters of the model. Deep learning, a subset of machine learning, focuses exclusively on neural networks.

The intrinsic parameters of a neural network, known as hyperparameters, are set before training, and include parameters such as the number of layers and neurons, epochs, and activation functions. Parameters of gradient descent and backpropagation algorithms \cite{rumelhart1986learning}, like batch size and learning rate, may also be hyperparameters. While some hyperparameters are predetermined, others are adjusted through tuning strategies.

ANNs are valued for their capacity to model large and complex datasets. The Universal Approximation Theorem asserts that an ANN with a single hidden layer and non-linear activation functions can model any nonlinear function \cite{hornik1990universal}, enhancing its utility for datasets with complex relationships. Even though an exhaustive review of ANNs is beyond the scope of this paper, great references exist in the literature \cite{nielsen2015neural, goodfellow2016deep}. For a basic introduction to their algorithms in the cosmological context, we recommend reading \cite{de2022observational}.

\subsection{Genetic algorithms}    
    \label{sec:ga}
Genetic algorithms are optimization techniques inspired by genetic population principles, treating each potential solution to an optimization problem as an individual. Initially, a genetic algorithm generates a population comprising multiple individuals within the search space. Across iterations or generations, the population evolves through operations like offspring, crossover, and mutation, progressively approaching the optimal solution of a target function. Genetic algorithms excel in addressing large-scale nonlinear and nonconvex optimization problems in challenging search scenarios \cite{gallagher1994genetic, sivanandam2008genetic}.

To apply genetic algorithms to a specific problem, one must select the objective function to optimize, delineate the search space, and specify the genetic parameters such as crossover, mutation, and elitism. Probability values for crossover and mutation operators are assigned, and a selection operator determines which individuals advance to the subsequent generation. Elitism, represented by a positive integer value, dictates the number of individuals guaranteed passage to the next generation. Overall, genetic algorithms initialize a population and iteratively modify individuals through the operators and the objective function, progressively approaching the optimal solution of the target function.

While this paper does not delve deeply into the mathematical principles underlying genetic algorithms, interested readers are directed to the following references \cite{reeves1997genetic, katoch2021review}, particularly for parameter estimation in cosmology \cite{Medel-Esquivel:2023nov}.

\section{Machine learning strategies}
\label{sec:strategies}
In this section, we outline our proposed method, which integrates machine learning techniques to implement neural networks and genetic algorithms within a nested sampling framework. \changes{Below we describe some deep learning techniques utilized in our training:}, elucidating their application:

\begin{itemize}
    \item  \textbf{Data scaling}. \changes{Since all samples within the parameter space are already scaled between 0 and 1 during nested sampling, no additional scaling is required for training the neural networks.}
    
    \item \textbf{Early stopping}. \changes{It is a regularization technique that monitors the performance of a model on a validation set during training and stops the training process when the performance on the validation set starts to degrade, indicating overfitting. It helps to prevent overfitting and choose the best weight configuration along the epochs of the training. By stopping the training process early, the generalization performance of the model can be improved, particularly when the training data is limited or noisy. We implement early stopping with a patience of 100 epochs to guarantee a minimum number of training epochs, given the smaller size of the dataset. However, our primary focus is on preserving the best-performing weights at the end of the training process.}
    
    \item \textbf{Dynamic learning rate}. There are popular strategies for dynamic learning rates. However, our dynamic learning rate is only adjusted during the nested sampling run and not during the training of a specific neural network. For each new training of the neural network, the learning rate decreases by half. \changes{However, during each individual ANN training session, the learning rate remains constant within the adaptive gradient descent algorithm called \texttt{Adam} \cite{kingma2014adam}.}
    
    \item \textbf{Hyperparameter tunning}. We have implemented the option of using genetic algorithms to find the architecture of the first trained neural network. For this purpose, we use the library \texttt{nnogada} \cite{gomez2023neural}. \changes{For simplicity in this work, we use genetic algorithms over 3 generations with a population size of 5 to explore combinations of batch size (4 or 8), number of layers (2 or 3), learning rate (0.0005 or 0.001), and number of neurons per layer (50 or 100). In a nested sampling execution, where we can train the neural networks multiple times, we use these small configurations. This approach yields better results compared to not tuning hyperparameters and is more effective than using a hyperparameter grid \cite{gomez2023neural}.}
\end{itemize}

We implemented our method inside of the code \texttt{SimpleMC} \cite{simplemc, aubourg2015}\footnote{The modified version of \texttt{SimpleMC} that includes our \texttt{neuralike} method is available at \url{https://github.com/igomezv/simplemc_tests}}, which uses the library \texttt{dynesty} \cite{speagle2020dynesty} for nested sampling algorithms. 
\changes{In all our neural network training, we use the mean squared error (MSE) as the loss function. If early stopping, with a patience of 100 epochs, does not stop the training, we select the configuration of weights that achieved the lowest MSE value.}

\subsection{\texttt{neuralike} method}
\label{sec:neuralike}

Neural networks are widely acclaimed for their formidable capabilities in handling extensive datasets. However, several studies have shown their effectiveness in modeling small datasets as well; even demonstrating that neural models can accommodate a total number of weights exceeding the number of sample data points \cite{ingrassia2005neural}. \changes{In addition, recent research} has focused on novel approaches by using neural networks with smaller datasets \cite{ng2015deep, pasini2015artificial, gomez2023neuralepjc}. \changes{While it is true that models with a large number of parameters can be prone to overfitting, this risk can be mitigated through the use of regularization techniques such as dropout and early stopping. In our approach, these techniques, combined with genetic algorithms for optimizing the network's architecture and hyperparameters, ensure that our models generalize well even when the number of parameters exceeds the number of data points.}

In nested sampling, as discussed in the previous section, there is a set of live points that maintain a constant number of elements. At a certain point in its execution, a new sample is extracted within a prior iso-likelihood or mass surface. Our goal is for the neural network to predict the likelihood of points within this prior volume. To do this, we train the neural network with only the current set of live points. These points, which typically are around hundreds or thousands, are sufficient to effectively train a neural network and have several advantages:
\begin{itemize}
    \item The relatively small dataset size implies that the neural network training process is not computationally intensive.
    \item By excluding points outside the current prior volume, we can potentially avoid inaccurate predictions in regions where points would be rejected based on the original likelihood. The points that the neural network learns efficiently are those within the prior volume, becasue they have a higher probability of acceptance according to the original likelihood.
    \item The quantity of elements within the training set remains constant. Whether the neural network starts its training at the beginning of sampling or at a later stage, the element count does not vary. As a result, the majority of neural network hyperparameters could stay consistent across different datasets.
\end{itemize}

\changes{The likelihood function in cosmological parameter estimation can be quite complex, often involving various types of observational data and intricate numerical operations, such as integrals, derivatives, or approximation methods for solving differential equations. To address this complexity, the idea is to replace the analytical likelihood function with a trained neural network. This substitution reduces the problem to a simple matrix multiplication, where the optimal weights, obtained during ANN training, are stored in a binary file. Consequently, the evaluation of the likelihood becomes significantly faster. This acceleration is particularly advantageous in a Bayesian inference process, where the likelihood function may need to be evaluated thousands or even millions of times, making the reduction in computational time highly beneficial.}
\\

Algorithm \ref{alg:pseudocode} provides an overview of our proposed methodology within a nested sampling execution. Concerning the neural network implementation, our primary focus is on the segment within the for loop. Once a predetermined number of samples have been reached, or when the flag \texttt{dlogz\_start} is activated, the ANN leverages the current live points for its training. The benefit of utilizing only the set of live points is twofold: firstly, it facilitates swift training, and secondly, it ensures that the ANN learns likelihood values strictly within the prior volume. This area is precisely where new samples should be located. 

\begin{algorithm*}
\caption{\footnotesize{Nested sampling with \texttt{neuralike}. \texttt{dlogz\_start} and \texttt{nsamples\_start} are the two ways to start \texttt{neuralike}, with a \texttt{dlogz} value (recommended) or given a specific number of generated samples. The \texttt{logl\_tolerance} parameter represents the neural network prediction tolerance required to be considered valid. \texttt{saved\_logl} denotes the log-likelihoods of the current live points, and \texttt{valid\_loss} determines the criterion for accepting or rejecting a neural network training. Any loss function values higher than \texttt{valid\_loss} will be rejected. The variable $logL$ represents the analytical log-likelihood function, while $\Like$ can either be $logL$ or $ANNmodel$, depending on the successful neural model.}}
\label{alg:pseudocode}
\begin{algorithmic}
\STATE
    using\_neuralike = False \\
    \uIf{livegenetic == True (optional)}
            {Define Pmut and Pcross\\
             Generate a population P with Nind individuals\\
             Evolve population through Ngen generations\\}
    \uElse{Generate Nlive live points}
                   
    \For{i in range{\rm (iteration)}}
    {
        \uIf{ ($dlogz < \texttt{dlogz\_start}$) OR ($\rm nsamples >= \rm \texttt{nsamples\_start}$) }
        {
        \uIf{i \% N == 0 AND using\_neuralike == False}
        {        
          Use nlive points as training dataset \\
          Optional: Use genetic algorithms with \texttt{nnogada} to choose the best architecture \\
          Use the best architecture to model the likelihood \\
          \uIf{loss function < \texttt{valid\_loss}}
                   {using\_neuralike = True\\
                   $\Like = \rm ANNmodel$\\}
          \uElse{continue with NS}
                   }
         \uIf{min(saved\_logl) - \texttt{logl\_tolerance} \ < neuralike < max(saved\_logl) + \texttt{logl\_tolerance}}
                  { 
                   {continue}
                    \uElse{like=logL; \\
                    using\_neuralike = False
                    }     
                      }      
        }
       }
\end{algorithmic}
\end{algorithm*}

\changes{It is important to note that the nested sampling process, including the selection of priors, typically uniform or Gaussian distributions, remains consistent with standard practices. Once the criteria for initiating ANN training are met, the live points are used to train the ANN. If the ANN’s performance metrics meet the required threshold, the analytical likelihood function is replaced by the ANN to save computational time. While this substitution does not alter the fundamental nested sampling process, it can significantly enhance efficiency by reducing computational overhead.}

\subsection{Using genetic algorithms}

We proposed genetic algorithms, like in our \texttt{nnogada} library \cite{gomez2023neural}, as an optional method to find the hyperparameter of the neural network as part of the workflow of \textit{neuralike}, as it can be noticed in the Algorithm \ref{alg:pseudocode}. In large parameter estimation processes, it is useful, despite the time required, to find the best neural network architecture.

On the other hand, we explored the first insight about the generation of the initial live points of a nested sampling process with genetic algorithms. It is analyzed in Appendix \ref{appendix}. Although we have incorporated the use of genetic algorithms in our code, the primary focus of this paper is on our \texttt{neuralike} method (Section \ref{sec:neuralike}). As such, further analysis of genetic algorithms in this context will be the subject of future research.


\section{Toy models}
\label{sec:toymodels}

As a first step in testing our method, we use some toy models as log-likelihood functions. These toy models only generate samplers within the Bayesian inference, without parameter estimation. However, it is useful to check the ability of the neural networks to learn, given a set of live points, the shape of these functions in runtime, and their respective values for the Bayesian evidence. We use the following toy models, with the mentioned hyperparameters:

\begin{itemize}
    \item A gaussian, \\ $f(x, y) = -\frac{1}{2}(x^2 + \frac{y^2}{2} - xy)$. Learning rate $5\times10^{-3}$, 100 epochs, batch size as 1.
    \item Eggbox function, \\ $f(x, y) = (2+\cos(\frac{x}{2.0})\cos(\frac{y}{2.0}))^{5.0}$. Learning rate $1\times10^{-4}$, 100 epochs, batch size as 1.
    \item Himmelblau's function, \\ $f(x,y)=(x^2+y-11)^2+(x+y^2-7)^2$. Learning rate $1\times10^{-4}$, 100 epochs, batch size as 1.
\end{itemize}

We have used some toy models as log-likelihood functions: Gaussian, egg-box, and Himmelblau. In Table \ref{tab:toymodels}, you can see the results of the Bayesian evidence calculation with and without our method for the three toy models, while in Figure \ref{tab:toymodels}, you can see the samples of the three functions, which at first glance are very similar. Based on these results, we can notice that for all these models, the speed of sampling using neural networks is slower than in the case of nested sampling alone; this is because the analytical functions are being evaluated directly without sampling from an unknown posterior distribution; nevertheless, these examples are very useful to verify the accuracy in calculating Bayesian evidence and sampling from the distribution. We can observe that both the log-Bayesian evidence and the graphs of the nested sampling process without and with neural networks are consistent; however, as Table \ref{tab:toymodels} shows, for more complex functions, we need a lower value of \texttt{dlogz\_start}, which means that we need to start learning the neural network at a later stage of nested sampling. Therefore, a lower \texttt{dlogz\_start} parameter is needed to be more accurate but slower, and it is precisely this parameter that regulates the speed-accuracy trade-off. 

\begin{table*}[]
    \centering
    \captionsetup{justification=raggedright, singlelinecheck=false, font=footnotesize}
    \begin{tabular}{|c|c|c|c|c|c|c|}
        \hline 
         Model & $\log Z$ & $\log Z$ \texttt{neuralike} & \texttt{dlogz\_{start}} & Valid loss & Samples & ANN samples \\
         \hline 
         Gaussian & $-2.13 \pm 0.05$ & $-2.16 \pm 0.05$  & 50 & 0.05 & 6774  & 6773\\
         Eggbox & $-235.83 \pm 0.11$ & $-235.82 \pm 0.11$  & 10 & 0.05 & 11794 & 3688 \\
         Himmelblau & $-5.59 \pm 0.09$ & $-5.64 \pm 0.09$ & 5 & 0.05 & 10253  &  4528 \\
         \hline 
    \end{tabular}
    \caption{Comparing Bayesian evidence for toy models with nested sampling alone and using \texttt{neuralike}. The column \texttt{dlogz\_start} indicates the dlogz value marking the start of neural network training; higher values suggest earlier integration of neural networks into Bayesian sampling. \textit{Valid loss} represents the threshold value of the loss function required for accepting a neural network as valid. The last two columns display the total number of samples generated through the nested sampling process and the subset produced by the trained neural networks.}
    \label{tab:toymodels}
\end{table*}

    \begin{figure*}[t!]
        \centering
        \captionsetup{justification=raggedright,singlelinecheck=false,font=footnotesize}
         \makebox[12cm][c]{
        \includegraphics[trim= 30mm 2mm 30mm 2mm, clip, width=3.8cm, height=4cm]{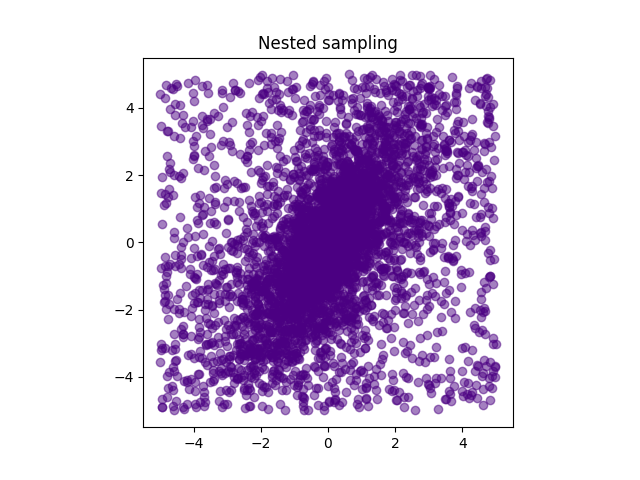}
        \includegraphics[trim= 30mm 2mm 30mm 2mm, clip, width=3.8cm, height=4cm]{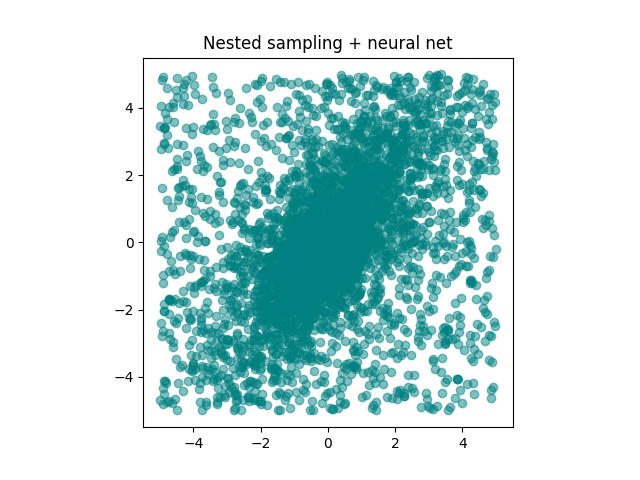}
        \includegraphics[trim= 30mm 2mm 25mm 2mm, clip, width=3.8cm, height=4cm]{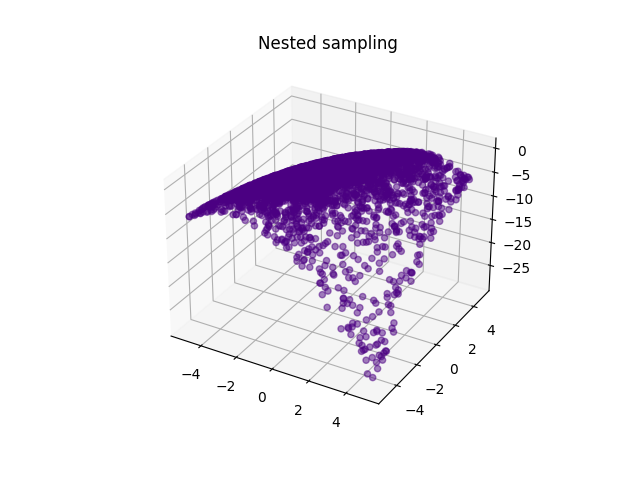}
        \includegraphics[trim= 30mm 2mm 25mm 2mm, clip, width=3.8cm, height=4cm]{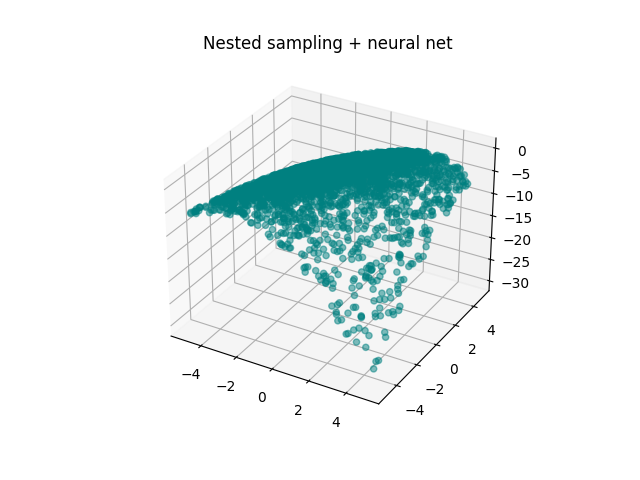}
        }
        \makebox[12cm][c]{
        \includegraphics[trim= 30mm 2mm 30mm 2mm, clip, width=3.8cm, height=4cm]{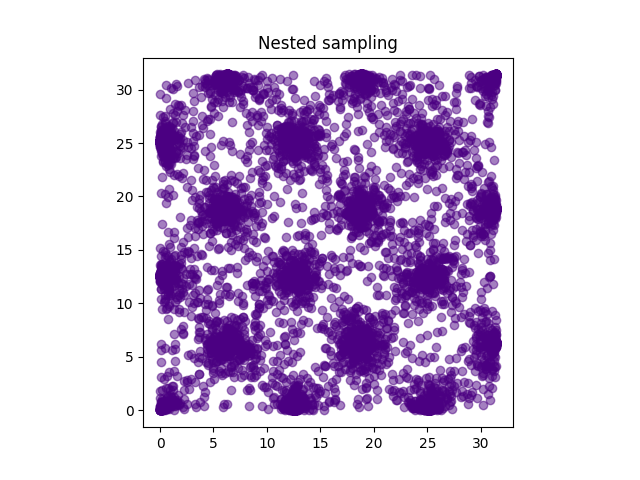}
        \includegraphics[trim= 30mm 2mm 30mm 2mm, clip, width=3.8cm, height=4cm]{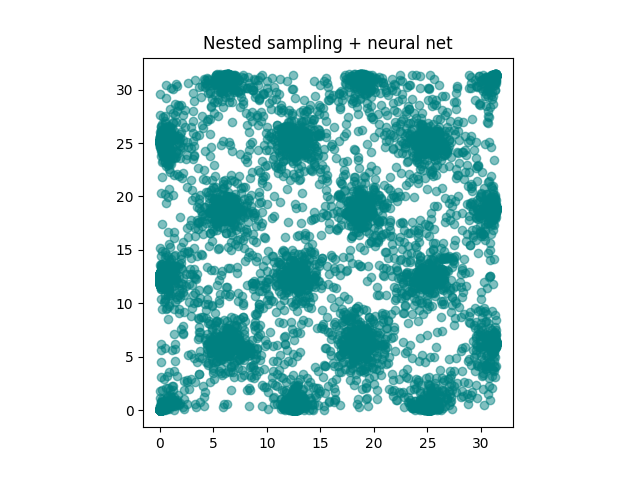}
        \includegraphics[trim= 30mm 2mm 25mm 2mm, clip, width=3.8cm, height=4cm]{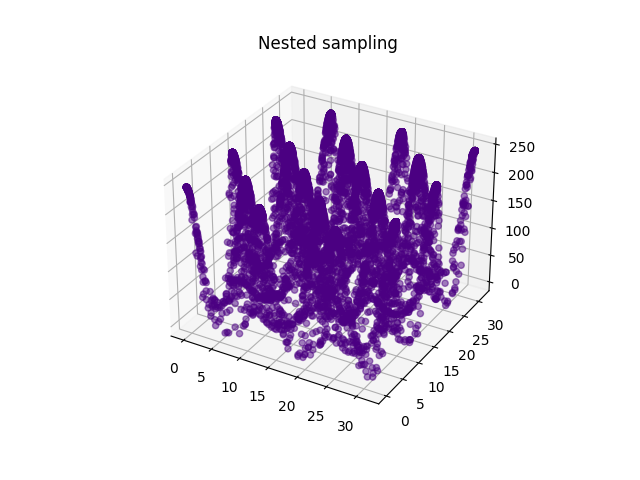}
        \includegraphics[trim= 30mm 2mm 25mm 2mm, clip, width=3.8cm, height=4cm]{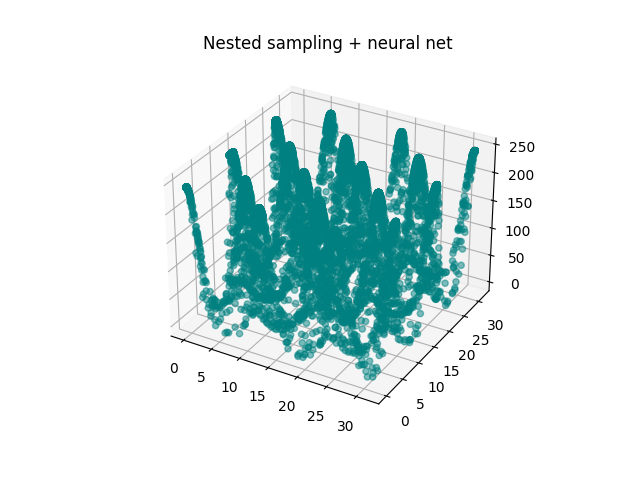}
        }
        \makebox[12cm][c]{
        \includegraphics[trim= 30mm 2mm 30mm 2mm, clip, width=3.8cm, height=4cm]{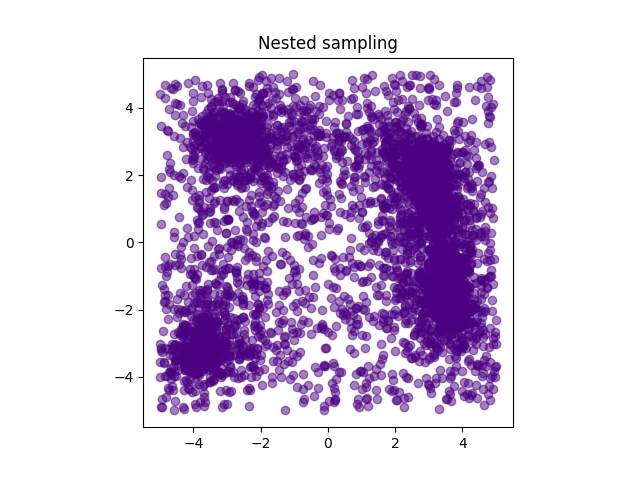}
        \includegraphics[trim= 30mm 2mm 30mm 2mm, clip, width=3.8cm, height=4cm]{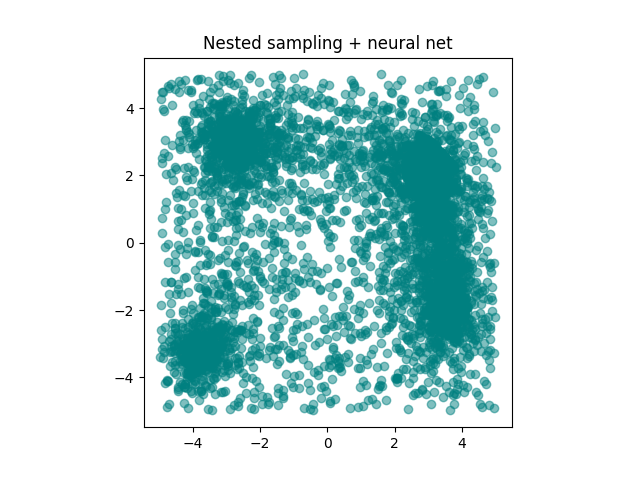}
        \includegraphics[trim= 30mm 2mm 25mm 2mm, clip, width=3.8cm, height=4cm]{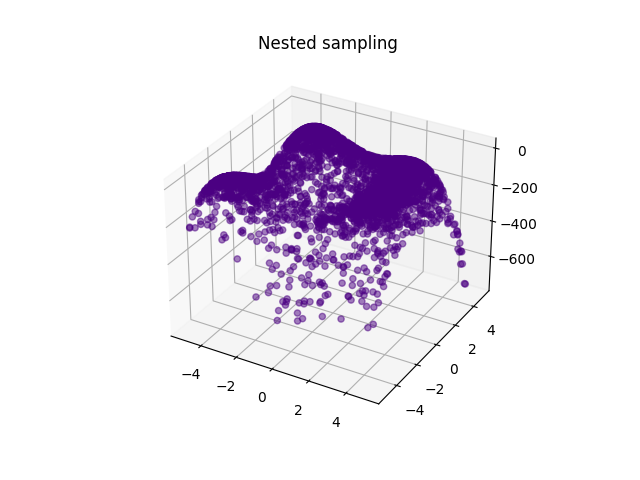}
        \includegraphics[trim= 30mm 2mm 25mm 2mm, clip, width=3.8cm, height=4cm]{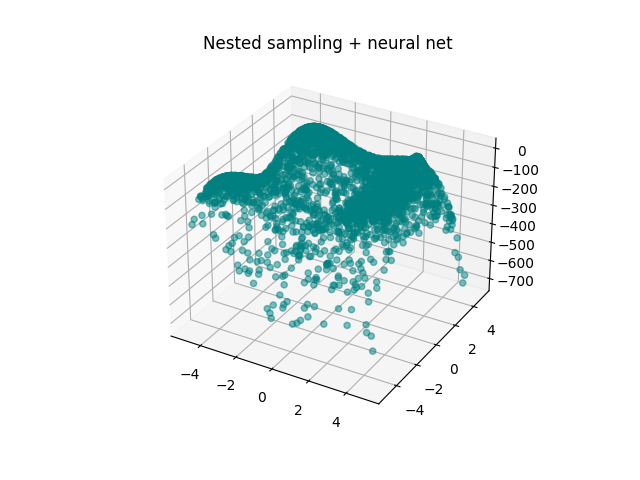}
        }
        \caption{Comparison of neural likelihoods versus original likelihoods using toy models. Using 1000 live points. }
        \label{fig:toymodels}
    \end{figure*}

\section{Cosmological parameter estimation} 
    \label{sec:cosmo}
 
     Assuming the geometric unit system where $ \hbar = c = 8\pi G = 1$, the Friedmann equation that describes the late-time dynamical evolution for a flat-$\Lambda$CDM model can be written as:
    \begin{equation}
        H(z)^2 = H_0^2\left [\Omega_{m,0}(1+z)^3 + (1-\Omega_{m,0})\right],
        \label{eq:hzlcdm}
    \end{equation} 
    where $H$ is the Hubble parameter and $\Omega_{m}$ is the matter density parameter;  subscript  0  attached  to  any  quantity  denotes its  present-day $(z=  0)$  value. In this case, the EoS for the dark energy is $w(z) = -1$.
     
    A step further to the standard model is to consider the dark energy being dynamic, where the evolution of its EoS is usually parameterized. A commonly used form of $w(z)$ is to take into account the next contribution of a Taylor expansion in terms of the scale factor  $w(a)= w_0 + (1-a)w_a$ or in terms of redshift $w(z) = w_0 + \frac{z}{1+z} w_a$ (CPL model \cite{chevallier2001accelerating, linder2003exploring}).
    The parameters $w_0$ and $w_a$ are real numbers such that at the present epoch  $w|_{z=0}=w_0$  and $dw/dz|_{z=0}=-w_a$; we recover $\Lambda$CDM when $w_0 = -1$ and $w_a=0$.
    Hence the Friedmann equation for the CPL parameterization turns out to be:
    \begin{equation}
    \begin{aligned}
        H(z)^2 = H_0^2 & [\Omega_{m,0}(1+z)^3 + \\ 
        & (1-\Omega_{m,0})(1+z)^{3(1+w_0+w_a)} e^{-\frac{3w_a z}{1+z}}].
        \label{eq:hzcpl}
    \end{aligned}
    \end{equation}
      
       
       In this work, we use cosmological datasets from Type-Ia Supernovae (SN), cosmic chronometers, growth rate measurements, baryon acoustic oscillations (BAO), and a point with Planck information. Following, we briefly describe them:
       
       \begin{itemize}
           \item \textbf{Type-Ia Supernovae}. We use the Pantheon SNeIa compilation, a dataset of 1048 Type Ia supernovae, with a covariance matrix of systematic errors $C_{sys} \in \mathbb{R}^{1048 \times 1048}$ \cite{scolnic2018complete}.
           
           \item \textbf{Cosmic chronometers}. Cosmic chronometers, also known as Hubble distance (HD) measurements, are galaxies that evolve slowly and allow direct measurements of the Hubble parameter $H(z)$. We use a compilation with $31$ data points collected over several years within redshifts between $0.09$ and $1.965$. \cite{jimenez2003constraints, simon2005constraints, stern2010cosmic, moresco2012new, zhang2014four, moresco2015raising, moresco20166, ratsimbazafy2017age}.
           
           \item \textbf{BAO}. We employ data from Baryon Acoustic Oscillation measurements (BAO) with redshifts $z<2.36$. They are from SDSS Main Galaxy Sample (MGS) \cite{ross2015clustering}, Six-Degree Field Galaxy Survey (6dFGS) \cite{beutler20116df}, SDSS DR12 Galaxy Consensus \cite{alam2017clustering}, BOSS DR14 quasars (eBOSS) \cite{ata2018clustering}, Ly-$\alpha$ DR14 cross-correlation \cite{blomqvist2019baryon} and Ly-$\alpha$ DR14 auto-correlation \cite{de2019baryon}, . 
           
           \item \textbf{Growth rate measurements}. We used an extended version of the Gold-2017 compilation available in \cite{sagredo2018internal}, which includes $22$ independent measurements of $f\sigma{_8}(z)$ with their statistical errors obtained from redshift space distortion measurements across various surveys.
           
           \item \textbf{Planck-15 information}. We also consider a compressed version of 
            Planck-15 information, where the Cosmic Microwave Background (CMB) is treated as a BAO experiment located at redshift $z= 1090$, measuring the angular scale of the sound horizon. For more details, see the Reference \cite{aubourg2015}.
       \end{itemize}

We executed three cases of parameter estimation to verify the performance of our method. We start with one thousand live points and a model with five free parameters; then, we increase the live points and free parameters, to test our method with higher dimensionality and with higher computational power demand (larger number of live points). The results are compared with a nested sampling run with the same data sets and the same configuration (live points, stopping criterion, etc.) but without ANN; this comparison aims to test the accuracy and speedup achieved by our \texttt{neuralike} method. For this comparison, we report the parameter estimation and Bayesian evidence obtained with and without our method and, in addition, we calculate the Wasserstein distances \cite{ramdas2017wasserstein} between the samples of the posterior nested sampling without and with \texttt{neuralike} for each free parameter considering their respective sampling weights.

In the results, a baseline neural network architecture was employed, configured with the following hyperparameters: 3 hidden layers, a batch size of 32, a learning rate of 0.001 (utilizing the Adam gradient descent algorithm for optimization), 500 epochs, and an early stopping patience of 200 epochs. In scenarios where multiple neural networks were required, the learning rate was reduced following the previously mentioned approach. As for evaluating the accuracy of the neural networks, we adopted a \texttt{valid\_loss} threshold of 0.05 for their training, and a \texttt{logl\_tolerance} of 0.05 for their predictions.

\subsection{Case 1}
\begin{figure*}[h!]
    \centering
    \captionsetup{justification=raggedright,singlelinecheck=false,font=footnotesize}
    \includegraphics[width=\textwidth]{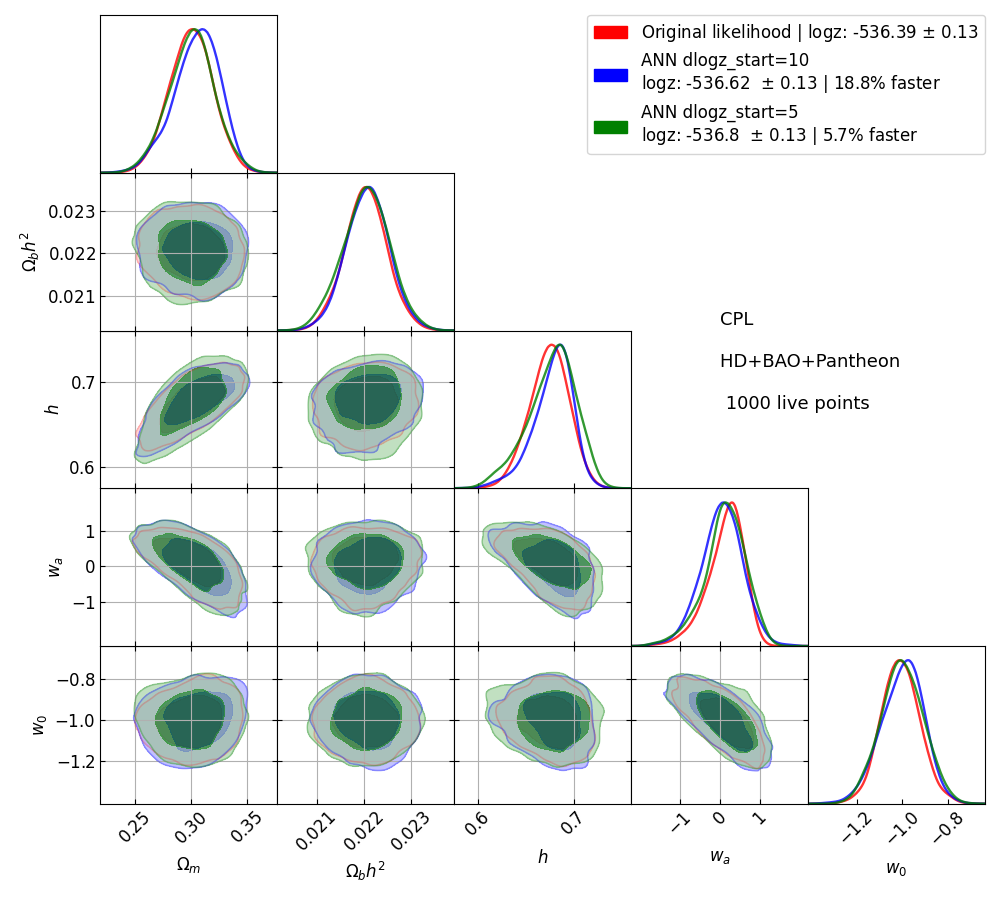}
    \caption{Case 1. Posterior plots for CPL Pantheon+HD+BAO with the proposed methods in this work.}
    \label{fig:posteriors1}
\end{figure*}

First, we perform the Bayesian inference for the CPL model using SNeIa from Pantheon compilation, with cosmic chronometers and BAO data. In this case, we consider only five free parameters: $\Omega_m$, $\Omega_bh^2$, $h$, $w_a$, and $w_0$. We use $1000$ live points. Figure \ref{fig:posteriors2} shows our results and we can notice that when $\texttt{dlogz\_start}=10$ the saved time is around $19\%$ and when $\texttt{dlogz\_start}=5$ it is around only $6\%$. According to Table \ref{tab:parameterestimation} both cases are in agreement with the $logZ$ value for nested sampling alone. If we check Table \ref{tab:estimation_diff} we can notice that, in general, the samples of the $\texttt{dlogz\_start} =5$ are more similar to the nested sampling posterior distributions; it also can be appreciated in the posterior plots shown in the Figure \ref{fig:posteriors1}. Although the case of $\texttt{dlogz\_start} =5$ saves less time than the case of $\texttt{dlogz\_start} = 10$, it gains in accuracy.

\subsection{Case 2}
\begin{figure*}[h!]
    \centering
    \captionsetup{justification=raggedright,singlelinecheck=false,font=footnotesize}
    \includegraphics[width=\textwidth]{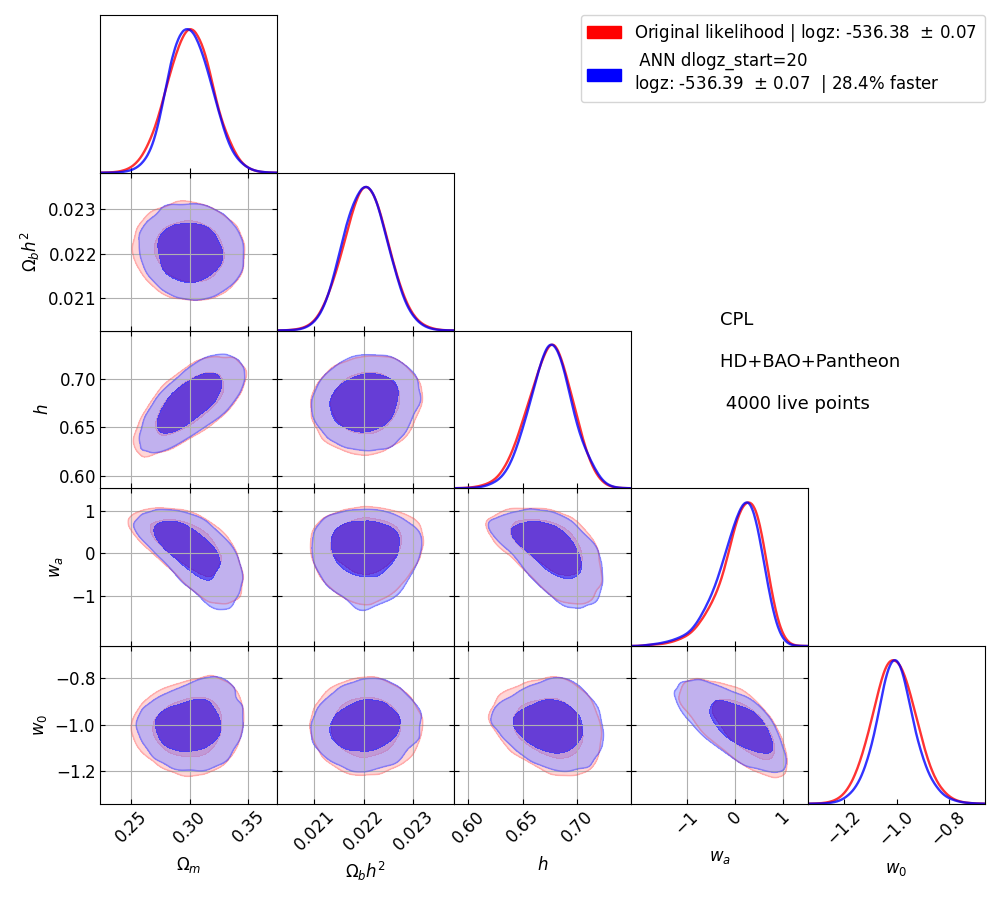}
    \caption{Case 2. Posterior plots for CPL using Pantheon+HD+BAO with the proposed methods in this work. We use 4000 live points.}
    \label{fig:posteriors2}
\end{figure*}

Secondly, we consider the same model, free parameters, and datasets as in Case 1. The difference in this second case is to analyze the behavior of our method with a larger number of live points. It has three new considerations: a) the training set for the neural network would be better because has a larger size, b) the number of operations in parallel for nested sampling is also larger, and c) we test whether the hypotheses based on a larger number of live points can obtain a better accuracy for the neural network earlier within the nested sampling process (i.e. in a higher value for \texttt{dlogz\_start}). Therefore, we increase the number of live points to 4000 and $\texttt{dlogz\_start = 20}$; the outputs are included in Figure \ref{fig:posteriors2} showing an excellent concordance for the Bayesian evidence values with our method, and speed-up around the $28.4\%$. Table \ref{tab:parameterestimation} contains the results of the Bayesian evidence, and it can be noticed that the uncertainty of this case is in better agreement with nested sampling than the two scenarios of Case 1. In addition, we can analyze Table \ref{tab:estimation_diff} and conclude that, effectively, its performance has a similar quality to Case 1 with $\texttt{dlogz\_start = 5}$; however because it uses a higher \texttt{dlogz\_start} value, the percentage of saved time is notorious.

\subsection{Case 3}
Lastly, we included more data: $f\sigma_8$ measurements and a point with Planck-15 information. To have more free parameters, eight in total, we consider contributions of the neutrino masses $\Sigma m_\nu$, growth rate $\sigma_8$, and curvature $\Omega_k$. In this case, we also used 4000 live points. With these new considerations, we aim to test our method in higher dimensions and to involve a more complex likelihood function that demands more computational power with each evaluation. We made several tests, but we include the corresponding to $\texttt{dlogz\_start}=5$, in which we obtain excellent results as can be noticed in Table \ref{tab:parameterestimation}. Due to the complexity of the likelihood, the full nested sampling process had to train three different neural networks, which allowed the use of erroneous predictions during sampling to be avoided. 

\begin{figure*}[h!]
    \centering
    \captionsetup{justification=raggedright,singlelinecheck=false,font=footnotesize}
    \includegraphics[width=\textwidth]{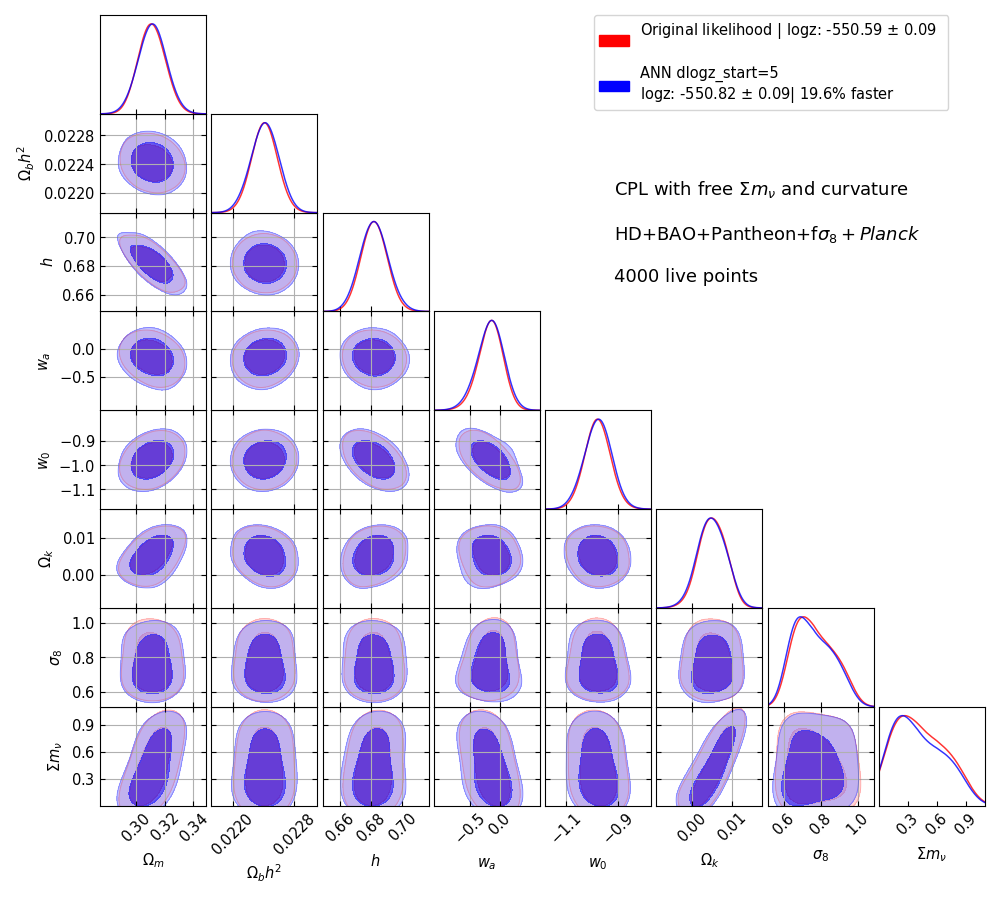}
    \caption{Case 3. 2D posterior plots for CPL with curvature using Pantheon+HD+BAO+f$\sigma_8$+Planck with the proposed methods in this work. Using 4000 points and considering 8 free parameters. In this case, because of the complexity, there were three neural networks trained before to substitute the likelihood function, however, the Bayesian inference process using our method was 19.6\% faster.}
    \label{fig:posteriors4}
\end{figure*}

We needed a lower value for the $\texttt{dlogz\_start}$ parameter due to the complexity of the model (given by the new free parameters); however, the saved time around of $19\%$ concerning the nested sampling alone is remarkable and the Wasserstein distance shown in Table \ref{tab:estimation_diff} indicates that the posterior distributions between the nested sampling with and without our method are similar, it also can be noticed in the posterior plots of the Figure \ref{fig:posteriors4}.

\begin{table*}[]
    \centering
    \captionsetup{justification=raggedright, singlelinecheck=false, font=footnotesize}
    \begin{tabular}{|c|c|c|c|c|c|c|}
        \hline 
          &\texttt{dlogz\_{start}} & $\log Z$ & $\log Z$ \texttt{neuralike} &  Samples & ANN samples & $\%$ saved time \\
         \hline 
        Case 1 & 10  & $ -536.39 \pm 0.13 $ & $ -536.62  \pm 0.13$   &   16001 & 9570   & 18.8\\
        Case 1 & 5  & $ -536.39 \pm 0.13 $ & $-536.8 \pm 0.13$  &  16297 &  7873  & 5.7\\
        Case 2 & 20  & $ -536.38 \pm 0.07$ & $-536.39  \pm 0.07$  &  65216 &  46147 & 28.4\\
        Case 3 & 5 & $ -550.59 \pm 0.09 $ & $-550.82 \pm 0.09$ & 95148 &  31923  & 19.6\\
         \hline 
    \end{tabular}
    \caption{Exploring Bayesian Inference with Nested Sampling and \texttt{neuralike}. The definitions of the columns are consistent with those in Table \ref{tab:toymodels}. Additionally, the \textit{$\%$ saved time} quantifies the speed-up achieved using our method.}

    \label{tab:parameterestimation}
\end{table*}

\begin{table*}[]
    \centering
    \captionsetup{justification=raggedright, singlelinecheck=false, font=footnotesize}
    \begin{tabular}{|c|c|c|c|c|c|c|c|c|}
        \hline 
          & $\Omega_m$ & $\Omega_bh^2$ & $h$ &  $w_0$ & $w_a$ & $\Omega_k$ & $\sigma_8$ & $\Sigma m_\nu$ \\
         \hline 
        Case 1a ($\texttt{dlogz\_start}=10$) & $0.00480$ & $0.00004$ & $0.00428$ & $0.01241$ & $0.11283$ & $-$ & $-$ & $-$\\
        Case 1b ($\texttt{dlogz\_start}=5$) & $0.00091$ & $0.00004$ & $0.00518$ & $0.00997$ & $0.05761$ & $-$ & $-$ & $-$\\
        Case 2 ($\texttt{dlogz\_start}=20$) & $0.00095$ & $0.00002$ & $0.00111$ & $0.00810$ & $0.06279$ & $-$ & $-$ & $-$\\
        Case 3 ($\texttt{dlogz\_start}=5$) & $0.00055$ & $0.00001$ & $0.00054$ & $0.00335$ & $0.01396$ & $0.00018$ & $0.00971$ & $0.01753$\\
         \hline 
    \end{tabular}
   \caption{Wasserstein distances \cite{ramdas2017wasserstein} between nested sampling posterior samples without and with \texttt{neuralike}, for each free parameter. The closer the value of this distance is to zero, the more similar are the distributions compared. This distance is implemented in \texttt{scipy} and takes into account the 1D posterior samples and their respective weights. Overall, parameters $\Omega_m$, $\Omega_bh^2$, and $h$ exhibit relatively small distances across all cases. However, in Case 1a, higher values of $w_0$ and $w_a$ distances are observed due to a higher \texttt{dlogz\_start} value and fewer data points used. On the other hand, Case 3 demonstrates smaller (better) distances, attributed to the utilization of more data and a lower \texttt{dlogz\_start} value.}

    \label{tab:estimation_diff}
\end{table*}

\section{Conclusions}
\label{sec:conclusions}

In this paper, we have introduced a novel method that incorporates a neural network trained on-the-fly to learn the likelihood function within a nested sampling process. The main objective is to avoid the time-consuming analytical likelihood function, thus increasing computational efficiency. We present the \texttt{dlogz\_start} parameter as a tool to handle the trade-off of accuracy and computational speed. In addition, we incorporate several deep learning techniques to minimize the risk of inaccurate neural network predictions.

To verify the effectiveness of our method, we employed several toy models, demonstrating their ability to replicate a probability distribution with remarkable accuracy in the nested sampling framework. Furthermore, in the cosmological parameter estimation, by performing a comparative analysis using the CPL cosmological model and various data sets, we highlighted the potential of our method to significantly improve the speed of nested sampling processes, without compromising the statistical reliability of the results. We found that, as the number of dimensions increased, our method produced a larger time reduction with a lower \texttt{dlogz\_start} value.

Despite commencing neural network training relatively late in the nested sampling process, the overall time reduction was notable, as evidenced in Table \ref{tab:parameterestimation}, showcasing reductions ranging from 6\% to 19\%. Potential errors in the neural network predictions were not found to be substantial because the training data set comprised the live points. As such, the likelihood predictions are not expected to deviate significantly from the actual prior volume, which enhances the credibility and robustness of our method and instills confidence in its application in nested sampling. In addition, our constant monitoring of the ANN prediction accuracy with the actual likelihood value allows us to be more confident in the results obtained, because if the criteria were not met, the analytical function would be used again and, after certain samples, another neural network would be retrained.

We also explore the potential utility of genetic algorithms in finding optimal neural network hyperparameters and in generating initial live points for nested sampling. Concerning the former, in scenarios where models are complex or high-dimensional, searching for an optimal architecture can be beneficial; however, our \texttt{neuralike} method allows this hyperparameter calibration to be optional so that hyperparameters can also be set by hand. Regarding the latter, we provide some insight into the potential advantages of using genetic algorithms to generate live points in Appendix A; however, future studies will address further research on this topic.

In this work, we only used observations from the late universe, as our \texttt{neuralike} method is integrated with the \texttt{SimpleMC} code that employs mainly background cosmology. However,  our method is easily applicable to the use of other types of observations, such as CMB data, an aspect we are currently working on.

We emphasize the importance of high accuracy in neural network predictions in observational cosmology since accurate parameter estimation is crucial for a robust physical interpretation of the results. In light of the machine learning strategies proposed in this paper, we can have greater confidence in the use of neural networks to accelerate nested sampling processes, without compromising the statistical quality of the results.

\section*{Acknowledgments}

IGV thanks the CONACYT postdoctoral grant, the ICF-UNAM support, and Will Handley for his invaluable advisory about nested sampling. JAV acknowledges the support provided by FOSEC SEP-CONACYT Investigaci\'on B\'asica A1-S-21925, FORDECYT-PRONACES-CONACYT 304001, and UNAM-DGAPA-PAPIIT IN117723. This worked was performed thanks to the help of the computational unit of the ICF-UNAM and the clusters Chalcatzingo and Teopanzolco.

\section*{Data Availability}

The implemented algorithm presented in this work is available in \url{https://github.com/igomezv/neuralike} and the original \texttt{SimpleMC} code in \url{https://github.com/javazquez/SimpleMC}, which contains the datasets used in this paper.

\newpage
\bibliographystyle{unsrt}
\bibliography{bibliography.bib}


\appendix

\section{Genetic algorithms as initial live points}
\label{appendix}

Previously, we mentioned that neural networks are good at interpolating, but not at extrapolating. Within the Bayesian inference process, we sample an indeterminate posterior probability distribution, whose shape remains unknown. Despite having some idea of the range of new samples in parameter space, we cannot definitively state that the highest likelihood point is already among the live points; this uncertainty may lead to inaccurate predictions for points close to the maximum likelihood point. In reference \cite{hogg2018data}, the authors propose the use of an optimizer to identify the optimal posterior probability sample, albeit at the expense of probabilism. The application of genetic algorithms to generate initial live points could be beneficial in cases where the Bayesian inference process must stop. In such circumstances, the partially generated posterior sampling aided by genetic algorithms will be more aligned with the maximum than a sampling generated without them. This alignment could facilitate a partial posterior sampling analysis. Although further investigation of this foray into genetic algorithms is needed, we have observed that when a small number of live points are used, and the initial live points are produced by a genetic algorithm, the stopping criterion is reached more quickly.

As the first insight into the genetic algorithms to generate the initial live points, we show some results about potential advantages in which genetic algorithms could help a nested sampling execution. In Table \ref{tab:genetic_appendix}, we can see some examples in which the use of GA to generate the first live points can reduce computational time without sacrificing the statistical results. However, it is worth noticing that we are using a low number of live points because this is the case in which we observed this advantage, when a higher number of live points is used, in general, NS alone is faster because have points in a sparse region of the search space and the use of GA cluster the points around the optimums losing exploration capacity. Nonetheless, there are possible scenarios in which there could be a low number of live points and in these cases, the incursion of GA to generate the initial sampling points could apport an advantage. This is part of a further study of the exploration in detail of this combination between GA with NS.

\begin{table}[]
    \centering
    \captionsetup{justification=raggedright, singlelinecheck=false, font=footnotesize}
    \scriptsize
    \begin{tabular}{|c|c|c|c|c|}
        \hline
         Model & eggbox &  eggbox & $\Lambda$CDM & $\Lambda$CDM \\
         \hline
         Sampler & NS  & NS+GA& NS & NS+GA \\
         \hline
         $\log Z$ & $-236.16 \pm 0.34$  & $-235.07 \pm 0.37$ & $-532.87 \pm 0.34$& $-532.70 \pm 0.33$\\ 
         $\Omega_m$ & $--$ & $--$ & $0.31 \pm 0.011$ & $0.31 \pm 0.011$\\
         $\Omega_bh^2$ & $--$ & $--$ & $ 0.02 \pm 0.0005$ & $0.02 \pm 0.0005$\\
         $h$ &$--$&$--$ &$0.683 \pm 0.009$& $0.683 \pm 0.009$  \\
         \tiny{\% saved time} & $--$ & 38  &$--$ & 23 \\
         \hline
    \end{tabular}
    \caption{Nested sampling for the eggbox toy model and $\Lambda$CDM using 100 live points. In the NS+GA cases, we generate the first live points through genetic algorithms with a probability of mutation equal to 0.5 and a probability of crossover of 0.8.}
    \label{tab:genetic_appendix}
\end{table}

\end{document}